# A CANDIDATE SUB-PARSEC SUPERMASSIVE BINARY BLACK HOLE SYSTEM WITH TWO SETS OF BROAD LINES

Todd A. Boroson & Tod R. Lauer
*National Optical Astronomy Observatory, Tucson, Arizona 85726, USA*

**The role of mergers in producing galaxies, together with the finding that most large galaxies harbour black holes in their nuclei[1], implies that binary supermassive black hole systems should be common. Here we identify a quasar, SDSS J153636.22+044127.0, as a plausible example of this phenomenon. This quasar shows *two* broad-line emission systems, separated in velocity by 3500 km s$^{-1}$. A third system of unresolved absorption lines has an intermediate velocity. These characteristics are unique among known quasars. We interpret this object as a binary system of two black holes, having masses of $10^{7.3}$ and $10^{8.9}$ solar masses separated by ~ 0.1 parsec with an orbital period of ~ 100 years.**

The present object is the first known with two broad-line systems and builds on the recent interest in the quasar SDSS J092712.65+294344.0, which shows two sets of *narrow* emission lines, only one of which has associated broad lines, separated by 2650 km s$^{-1}$ from each other[3,4,5,6,7], as well as additional emission and absorption lines at an intermediate redshift[7]. While these characteristics might be caused by a binary black hole system, they might also be due to chance superposition of active objects or a colliding system analogous to NGC 1275[6].

We have developed a principal components analysis technique that identifies objects having spectral characteristics inconsistent with the ensemble of quasar spectra in the SDSS archive. We have applied this procedure to the rest spectral region 4000-5700 Å of ~17,500 quasars. This sample comprises all quasars having redshifts less than 0.70 from the SDSS Quasar Catalog, fifth release[8] (SDSSQ5 catalog), as well as all additional objects classified as z<0.70 quasars observed for the final release (DR7). Of the 17,500 objects in the entire sample, only two objects have multiple redshift systems, SDSS J092712.65+294344.0, described above, and SDSS J153636.22+044127.0 (J1536+0441).

J1536+0441 has SDSS point-source (PSF) *ugriz* magnitudes of 17.72, 17.24, 16.97, 16.80, and 16.33. The image appears stellar in SDSS images. For a redshift of 0.388 and a flat cosmology with $H_0$ = 71 km s$^{-1}$ Mpc$^{-1}$ and $\Omega_m$ = 0.27, the absolute *i*-band magnitude is $M_i$=-24.83. This is at the high end of the luminosity distribution for the SDSS quasar sample at this redshift. The quasar has 2MASS JHK magnitudes of 15.46, 14.85, and 14.10. It was also detected by ROSAT, with a count rate of 0.03124 s$^{-1}$. This is typical for SDSS quasars at this redshift. The quasar is not in the FIRST or NVSS catalogues.

The spectrum of J1536+0441 shows two broad-line emission systems and one system of narrow absorption lines. The higher redshift "r-system" at z = 0.3889 shows broad Balmer lines (Hα, Hβ, and Hγ) and the usual narrow lines seen in low redshift quasars. The lower redshift "b-system" at z = 0.3727, shows broad Balmer lines (Hα through Hδ) and broad Fe II emission, seen most strongly around 3000 Å in the rest frame.  A strong narrow absorption-line  "a-system" is also present, including 6 unresolved resonance lines, at z =0.38783, which, in the quasar rest frame, is 240 km s$^{-1}$ less than that of the r-system and 3300 km s$^{-1}$ greater than that of the b-system.  The full SDSS spectrum is shown in Figure 1.  The lines are listed in Table 1.

The r-system shows the typical features of a low-redshift quasar.  The strengths and widths of the forbidden lines are normal.  The somewhat larger width of the high ionization forbidden lines ([Ne V]) compared to that of the low ionization forbidden lines ([O II]) is not unusual[9,10].  The Balmer line profiles, to the extent that they can be separated from the b-system, look normal.

The b-system has no narrow or forbidden-line emission, making it extremely unusual.  An upper limit on the equivalent width of  its [O III] λ5007 line is  ~ 0.5 Å, about 2% of the measured strength of the r-system [O III] λ5007 line.  While there are a few quasars known with no detectable [O III] lines, they are exclusively IR-luminous objects that have extremely strong optical Fe II emission[11].  Thus, they are quite rare and show features not seen in this object.

UV Fe II emission is unambiguously indicated by the "notch" between about 3950 and 4300 Å. To determine to which system the Fe II emission should be attributed, the spectral region between 4200 and 4450 Å was cross-correlated with a composite quasar spectrum[12] (Figure 2).  It is clear that the Fe II belongs to the b-system, with a derived redshift of 0.375.  While this is slightly higher (500 km s$^{-1}$) than that of the b-system, Fe II emission in quasars is typically redshifted by 400 km s$^{-1}$ relative to the Hβ line[13].

The absorption line system is also atypical.  The lines are unresolved in the SDSS spectra, and are almost certainly due to neutral or low-ionization gas in the line of sight rather than stars in the host galaxy.  There is no evidence of stellar absorption features.  Absorption such as this is very unusual in low-redshift quasars.  Only 6 of the 1000 high signal-to-noise spectra of SDSS quasars that were used for constructing the eigenspectra show Na I D absorption at the strength seen in J1536+0441, and of these, three are clearly cases of self-absorption from dense material associated with the nucleus.  It is possible that this absorption arises in a gas cloud unrelated to the presence of activity in the host galaxy – even in some other object in the line of sight.  It is also possible that this absorption is seen because of the presence of the second continuum source, presumably the b-system nucleus, seen through a part of the r-system that, normally, a line of sight to the nucleus would not pass.

If the two broad-line redshifts represent broad-line emission regions around separate supermassive black holes in a single galaxy, then the simplest model would be one in which they are in a gravitationally bound system.  We estimate the masses of the two black holes using the Hβ FWHM values and assuming that a fraction, proportional to the mass, of the continuum luminosity at λ5100 comes from each quasar.  Adopting a standard calibration[14], we derive masses of $10^{8.9}$ and $10^{7.3}$ solar masses for the red and blue systems respectively.

Derivation of the full orbital velocity, $V$, radius, $R$, and period, $T$, of the binary system depends on unknown geometrical factors. If a random circular orbit is assumed, the mean angle between the line-of-sight and the radial vector between the two black holes is 60°. Random orientation of the velocity vector implies an additional mean projection angle of 45°. The projected velocity is then $\approx 0.61V$, which gives $V \sim 6 \times 10^3$ km s$^{-1}$. This implies $R \sim 3 \times 10^{17}$ cm and $T \sim 100$ years. In contrast, upper limits of $T \sim 500$ years and $R \sim 9 \times 10^{17}$ cm are derived if no projection factors are assumed. In either case, the separation is approximately the size of the broad line region and very much smaller than the narrow line region. Thus, in this model, the two black holes are orbiting with their broad line regions well within a single narrow line region. It is notable that the derived characteristics of this system are similar to those proposed for OJ 287 based on its long term photometric variations[15,16].

For fixed black hole masses, the decay time of the binary due to gravitational radiation, $t_D \sim V^{-8}$, and is thus extremely sensitive to the assumed projection factors. For no projection corrections, $t_D \sim 3 \times 10^{11}$ years, while $t_D \sim 7 \times 10^9$ years under the model above[17]. This timescale is interesting, as it implies that the binary has evolved past the "final parsec" scale at which decay due to energy exchange with stars becomes inefficient, but where gravitational radiation decay remains too weak to carry the evolution further[18]. Theoretical studies of the effects of gas dynamical friction indicate that the timescale for that process to cause the orbit to decay is even much longer for such massive black holes[19,20], though this is an area of ongoing study.

It is intriguing that the two black hole masses derived would put the systemic velocity closer to the r-system, consistent with the idea that the narrow emission lines and absorption lines are more likely associated with the host galaxy. Note that the masses are highly uncertain both because of the difficulty of measuring the width of the blended lines and the uncertainty about how to split continuum flux between the two objects. This interpretation is further complicated by the fact that the narrow forbidden emission lines agree precisely with the redshift of the r-system.

A second possibility is that these are two separate quasars that are by chance seen in the same line of sight. Integrating the SDSS quasar luminosity function[21] yields a probability of $1.8 \times 10^{-7}$ of finding a second quasar within a volume centred on a first quasar corresponding to a conservative one-arcsecond radius circle on the sky and a redshift range of $\pm 10{,}000$ km s$^{-1}$. Multiplying this by the 17,500 objects in our sample gives a probability of $3.2 \times 10^{-3}$ of one chance occurrence. This calculation does not account for the possibility of an enhanced density due to the presence of a cluster of galaxies; two galaxies near the quasar on the SDSS image have photometric redshifts close to that of the quasar. The strongest, though admittedly *a posteriori,* argument against the chance superposition hypothesis is the unique lack of narrow lines in the b-system.

The most obvious interpretation of the presence of absorption at a redshift close to the r-system is that the b-system is background to the r-system. This would be inconsistent with an explanation involving ejection of one of the black holes, though it would be consistent with an infall interpretation, analogous with NGC 1275[22]. In this case, however, the two broad-line systems must represent two active nuclei.

New observations can test the binary black hole hypothesis. A spectrum with higher signal-to-noise may allow the detection of stellar absorption features from the host galaxy, as well as setting better limits on any narrow emission lines associated with the b-system. Detection of such features at both b- and r-system redshifts would provide compelling evidence that this is a chance superposition or colliding pair. High spatial resolution imaging could also rule out the close binary hypothesis. Monitoring over several years could reveal changes due to the orbital motion of the system. Our simple projection model predicts a velocity change in the b-system of $\sim 10^2$ km s$^{-1}$ in a single year.

**Acknowledgements**  NOAO is operated by the Association of Universities for Research in Astronomy (AURA), Inc. under cooperative agreement with the National Science Foundation.

This paper has used data from the Sloan Digital Sky Survey (SDSS) archive; the Two Micron All Sky Survey (2MASS) archive; the Röntgen Satellite (ROSAT) archive at the High Energy Astrophysics Science Archive Research Center (HEASARC), provided by NASA's Goddard Space Flight Center, the Faint Images of the Radio Sky at Twenty Centimeters (FIRST) survey, and the NRAO VLA Sky Survey (NVSS).

Funding for the SDSS and SDSS-II has been provided by the Alfred P. Sloan Foundation, the Participating Institutions, the National Science Foundation, the U.S. Department of Energy, the National Aeronautics and Space Administration, the Japanese Monbukagakusho, the Max Planck Society, and the Higher Education Funding Council for England. The SDSS Web Site is http://www.sdss.org/. The Two Micron All Sky Survey is a joint project of the University of Massachusetts and the Infrared Processing and Analysis Center/California Institute of Technology, funded by the National Aeronautics and Space Administration and the National Science Foundation.

**Author information**  Reprints and permissions information is available at www.nature.com/reprints. Correspondence and requests for materials should be addressed to T.B. (tyb@noao.edu).

TABLE 1

**SPECTRAL LINE PROPERTIES OF SDSS J1536+0441**

| System | Line | Z | FWHM (km s$^{-1}$) | EW (Å) |
|---|---|---|---|---|
| r-system (emission) <br> (<z> = 0.3889) | Hα | 0.38890 | - | - |
| | [O III] 5007 | 0.38889 | 550 | 24 |
| | [O III] 4959 | 0.38889 | 530 | 8 |
| | Hβ | 0.38894 | 6000: | 43 |
| | [O III] 4363 | 0.38820 | - | 0.8 |
| | Hγ | 0.38914 | - | 5.1 |
| | [Ne III] 3968 | 0.38941 | 340: | 0.5 |
| | He I 3889 | 0.38960 | 920 | 1.5 |
| | [Ne III] 3869 | 0.38872 | 680 | 2.1 |
| | [O II] 3727 | 0.38908 | - | 1.8 |
| | [Ne V] 3426 | 0.38912 | 1030 | 0.9 |
| b-system (emission) <br> (<z> = 0.3727) | Hα | 0.37303 | 2100: | 165 |
| | Hβ | 0.37247 | 2400: | 35 |
| | Hγ | 0.37253 | - | 8: |
| | Hδ | 0.3717: | - | 0.4: |
| a-system (absorption) <br> (<z> = 0.38783) | Na D 5896 | 0.38782 | unresolved | 0.8 |
| | Na D 5890 | 0.38794 | unresolved | 1.2 |
| | Ca II K 3934 | 0.38776 | unresolved | 0.4 |
| | Mg I 2852 | 0.38784 | unresolved | 0.9 |
| | Mg II 2796 | 0.38781 | unresolved | 2.1 |
| | Mg II 2803 | 0.38779 | unresolved | 2.5 |

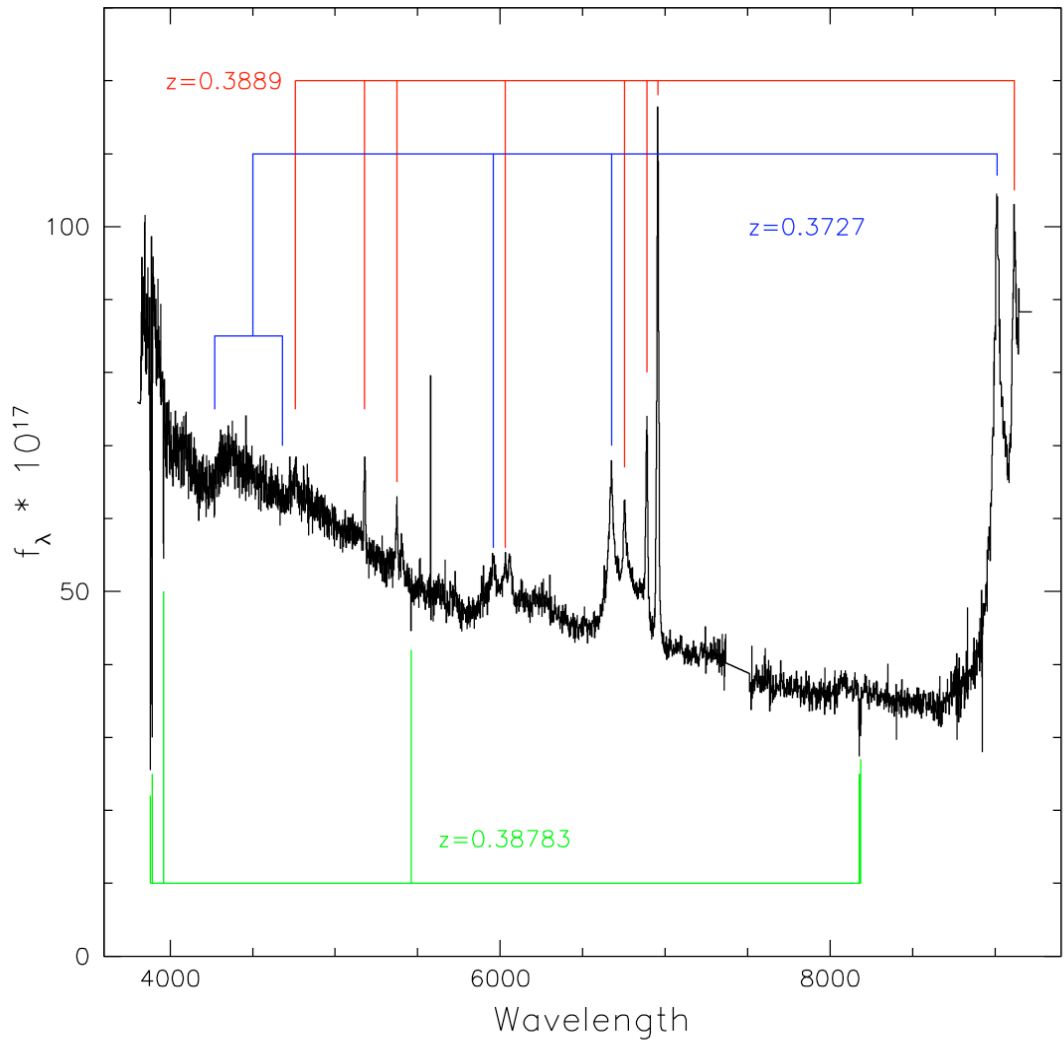

Figure 1 – **The observed spectrum from the SDSS archive of the quasar SDSS J153636.22+044127.0** The three redshift systems discussed in the text are indicated with identified features marked. The r-system, at z=0.3889, shows typical broad and narrow lines seen in low-redshift quasars, including the Balmer lines and the strong forbidden lines of [O II], [O III], [Ne III], and [Ne V]. The b-system, at z=0.3727 shows only broad Balmer lines and UV Fe II emission. The a-system shows 6 unresolved absorption lines: the Mg II doublet ($\lambda\lambda$2796, 2803), the Mg I $\lambda$2852 line, the Ca II K line ($\lambda$3934), and the Na D doublet ($\lambda\lambda$5891,5897). The strong unmarked emission feature is an artifact from poor subtraction of the nightsky line at $\lambda$ 5577.

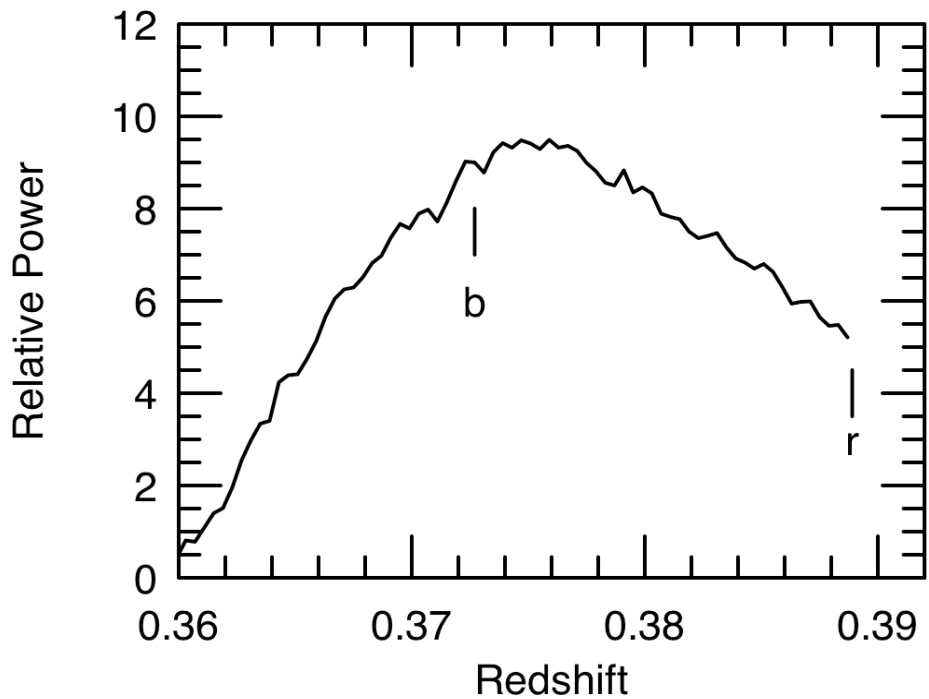

Figure 2 – **Cross correlation between the 4200-4450 Å region of the spectrum of J1536+0441 and the corresponding region of a composite quasar spectrum**[13]. The redshifts of the b- and r- systems are marked. The peak indicates the presence of UV Fe II at, or slightly above, the redshift of the b-system. Note that the cross correlation is quite inconsistent with the Fe II emission being at the redshift of the r-system (z=0.3889).